\documentclass[sigconf]{acmart}

\AtBeginDocument{%
  }

\usepackage{listings}
\usepackage{xcolor}
\usepackage{booktabs}
\usepackage{xspace}

\usepackage{subcaption}
\usepackage{graphicx}
\usepackage{adjustbox}
\usepackage{wrapfig}
\usepackage{tabularx}
\usepackage{booktabs}
\usepackage{ragged2e}  
\usepackage{multirow}
\usepackage{fancyvrb}
\usepackage{float}
\newfloat{listing}{htbp}{lst}
\floatname{listing}{Listing}

\usepackage{listings}
\usepackage{xcolor}
\usepackage{textcomp}

\definecolor{codepurple}{rgb}{0.58,0,0.82}
\definecolor{codered}{rgb}{0.8,0,0}
\definecolor{codegreen}{rgb}{0,0.6,0}
\definecolor{codewhite}{rgb}{1,1,1}

\newcommand{\CodeIn}[1]{\begin{small}\texttt{#1}\end{small}}

\newcommand{\condprob}[2]{\CodeIn{P(#1|#2)}}

\copyrightyear{2026}
\acmYear{2026}
\setcopyright{cc}
\setcctype{by}
\acmConference[FSE Companion '26]{34th ACM Joint European Software Engineering Conference and Symposium on the Foundations of Software Engineering}{July 05--09, 2026}{Montreal, QC, Canada}
\acmBooktitle{34th ACM Joint European Software Engineering Conference and Symposium on the Foundations of Software Engineering (FSE Companion '26), July 05--09, 2026, Montreal, QC, Canada}
\acmDOI{10.1145/3803437.3805237}
\acmISBN{/2026/07}

\begin{document}

\title{Dynamic Cogeneration of Bug Reproduction Test in \\Agentic Program Repair}

\newcommand{\Google}{
\affiliation{%
  \institution{Google}
  \country{}
}
}

\author{Runxiang Cheng}
\Google
\email{chengsam@google.com}

\author{Michele Tufano}
\Google
\email{tufanomichele@google.com}

\author{Jos\'{e} Cambronero}
\Google
\email{jcambronero@google.com}

\author{Renyao Wei}
\Google
\email{renyaow@google.com}

\author{Sherry Shi}
\Google
\email{sherryyshi@google.com}

\author{Grant Uy}
\Google
\email{grantuy@google.com}

\author{Pat Rondon}
\Google
\email{rondon@google.com}

\author{Franjo Ivan\v{c}i\'{c}}
\Google
\email{ivancic@google.com}

\renewcommand{\shortauthors}{R. Cheng, M. Tufano, J. Cambronero, R. Wei, S. Shi, G. Uy, P. Rondon, and F. Ivan\v{c}i\'{c}}
\renewcommand{\shorttitle}{Dynamic Cogeneration of Bug Reproduction Test in Agentic Program Repair}

\begin{abstract}
Bug Reproduction Tests (BRTs) have been used in many Automated Program Repair (APR) systems, primarily for validating fixes and aiding fix generation.
In practice, when developers submit a patch, they often implement the BRT alongside the fix. Our experience deploying agentic APR reveals that developers desire a BRT within AI-generated patches to increase their confidence.
However, canonical APR systems tend to generate BRTs and fixes separately, and focus on producing only the fix in the final patch.
In this paper, we study agentic APR in the context of cogeneration, where the APR agent is instructed to generate both a fix and a BRT in the same patch.
We evaluate the effectiveness of different cogeneration strategies on 120 human-reported bugs at Google and characterize different cogeneration strategies by their influence on APR agent behavior.
We develop and evaluate patch selectors that account for test change to select patches with plausible fixes (and plausible BRTs). 
Finally, we analyze the root causes of failed cogeneration trajectories. 
We show that cogeneration allows the APR agent to generate BRTs for at least as many bugs as a dedicated BRT agent, without compromising the generation rate of plausible fixes, thereby reducing engineering effort in maintaining and coordinating separate generation pipelines for fix and BRT at scale. 
\end{abstract}


  
\maketitle
\section{Introduction}
\label{sec:intro}

A Bug Reproduction Test (BRT) is a test that fails due to the presence of the bug and passes once the bug is fixed. 
In bug fixing, a BRT is often written to help validate whether a fix resolves the bug, and added to the codebase along with the fix to detect regressions in the future.
As more agentic Automated Program Repair (APR) systems emerge, BRT generation has also been integrated into some of these systems.
For example, some agentic APR systems have dedicated components that generate BRTs, and use them to assess generated fixes~\cite{kang2025autocodesherpa,xia2025demystifying,ruan2025specrover,arora2024masai,li2025infcode}.
Some APR agents are instructed to write bug reproduction scripts for fix verification while generating fixes~\cite{yang2024swe,gao2025trae}. 
Recently, LLM-based systems and agents have also been developed specifically to generate BRTs~\cite{khatib2025assertflip,ahmed2025heterogeneous,kitsios2025automated,nashid2025issue2test,ahmed2025otter,mundler2024swt,cheng2025agentic}. 

In the industry setting, when developers submit a bug-fixing code patch, the BRT is often implemented concurrently with the fix and included in the same patch as the fix.
In our live deployment of an agentic APR system, developers would also prefer a BRT to be present in the AI-generated fix patch to increase the reviewers' confidence in the patch.
However, existing LLM-based APR systems return a final patch with only the fix while discarding the generated BRT that was used to derive the fix~\cite{kang2025autocodesherpa,xia2025demystifying,ruan2025specrover,arora2024masai,li2025infcode,yang2024swe,gao2025trae}, and most of them further separate the fix and the BRT generation pipelines~\cite{kang2025autocodesherpa,xia2025demystifying,ruan2025specrover,arora2024masai,li2025infcode}.
Given the intertwined nature of bug fixing and bug reproduction, we hypothesize that merging both generation tasks in agentic APR could be beneficial.

This integration of fix and test generation invites a re-evaluation of traditional software engineering practices within the agentic domain.
In the realm of human software development, the relationship between testing and coding has been rigorously scrutinized~\cite{fucci2016dissection}.
Empirical studies of developer behavior reveal that developers predominantly write tests only after the implementation is complete, treating testing as a subsequent verification phase rather than a design driver~\cite{beller2015and,munir2014experimental}.
Conversely, established best practices often advocate for Test-Driven Development (TDD)~\cite{beck2003test} not merely as a validation technique, but as ``cognitive scaffolding''---a disciplined workflow to reduce cognitive load and guide problem-solving through incremental steps~\cite{beck2003test, janzen2005test}.
While this scaffolding is theorized to aid human cognition, its transferability to AI agents remains unverified.
The core tension lies between the disciplined structure of these workflows and the stochastic, exploratory nature of AI models.

To that end, we present the first study of end-to-end agentic APR that dynamically generates and returns to the user both fix and BRT in the same trajectory. We empirically evaluate different cogeneration strategies inspired by \textit{human developer workflows} to systematically investigate their efficacy within the domain of agentic program repair. Specifically, we employ: Test-Driven Development (TDD)---writing the test before the fix, Test-Last Development (TLD)---writing the test after the fix, and Freeform---where the implementation order is at the discretion of the agent.

This ``cogeneration'' setup further provides several benefits from the production standpoint.
First, developers prefer to see BRTs when reviewing bug fixes---cogeneration organically allows the APR agent to produce patches with coherent (fix, BRT) implementations when possible.
Second, it enables the agent to reuse context shared between fix and BRT generation tasks (e.g., root cause analysis), while avoiding the overhead of developing and coordinating separate generation components.
Third, the agent can alternately revise its fix/test implementations, and its performance scales transparently with the LLM’s reasoning and coding ability.

We compare these three cogeneration strategies (TDD, TLD, and Freeform) against 2 baselines: a Fix-only agent---writing only the fix, and a BRT-only agent---writing only the BRT.
We use execution-based metrics (e.g., pass@k~\cite{chen2021evaluating,jimenez2023swe,austin2021program}, plausibleBRT@k) to measure effectiveness on 120 human-reported bugs across 6 programming languages from  Google's Issue Tracking System (GITS).
We find that all cogeneration strategies generate plausible fixes and BRTs for at least as many bugs as Fix-only and BRT-only, respectively, with Freeform cogeneration being the best.

We characterize different cogeneration strategies based on their impact on the agent’s behaviors.
We analyze how they affect the occurrence of important bug fixing behaviors (e.g., fault localization success, test modification) and common action transitions of the agent.
We also define and measure cogeneration-specific success rates to characterize the expected efficacy of different cogeneration strategies.
Our results largely confirm their expected behaviors and efficacy.
We also find that the agent often defaults to following TLD when not enforced to follow any workflow (i.e., Freeform).

To adapt patch selection to the context of cogeneration, we implement and evaluate test-aware patch selectors that can leverage the BRT information from the cogenerated patches to group and select patches accordingly.
We evaluate them against the default, test-unaware patch selector.
We also assess how different BRT quality signals in a cogenerated patch can provide signals of the quality of the corresponding fix.
Our goal is to prioritize selecting patches with both a plausible fix and BRT over patches with only a plausible fix.
Our results show that the best test-aware patch selector reaches a 0.16/0.71 precision/recall on patches with plausible fix and BRT, compared to the default selector (0.08/0.57); their precision/recall on patches with only a plausible fix are similar (0.30/0.81 vs. 0.31/0.76).

Lastly, we perform a qualitative analysis on the common failures and respective root causes of finished cogeneration trajectories that produced implausible outcomes, and discuss possible mitigation.
The top issues are (1) the cogenerated patch has no test, because the agent considered the BRTs to be temporary changes and cleaned them up before finishing, (2) the agent exhausted steps in pursuing a wrong debugging hypothesis,  (3) the agent implements a fix overfitted to its BRT (or vice versa), thereby failing  the oracle patch.

To summarize, this paper makes the following contributions:
\begin{itemize}
    \item An efficacy study of an APR agent when cogenerating fix and BRT together versus when generating fix or BRT alone.
    \item A characterization study on the impact of different cogeneration strategies on APR agent behaviors.
    \item An adaptation of patch selection for cogeneration, by implementing and evaluating test-aware patch selectors for selecting cogenerated patches that have both fix and BRT.
    \item A qualitative analysis of common failures and corresponding root causes on cogeneration trajectories that finished with suboptimal outcomes.
\end{itemize}

\section{Background}
\label{sec:background}

We present the background context on our agentic APR system and motivation for studying cogeneration of BRTs.

\subsection{Our Agentic APR System}
\label{sec:background:agent}

Our agentic APR system, named Passerine, consists of three main components: bug abstention, patch generation, and patch validation \& selection~\cite{rondon2025evaluating,cambronero2025abstain}.
Passerine is deployed to fix internal bugs and failures reported at Google.

Abstention omits bugs that are likely unfit for patch generation (e.g., those with vague description or require no code change), and patch validation \& selection filters out invalid generated patches (e.g., those that failed regression tests) and selects at most one patch per bug to be sent to human developers for review~\cite{cambronero2025abstain}.
A patch will be committed to Google’s monorepo and the corresponding bug will be marked as fixed once it passes human review.

The patch generator is an LLM-driven ReAct-style~\cite{yao2022react} code generation agent for APR~\cite{rondon2025evaluating}.
To start a repair trajectory for a bug, the APR agent is provided with system instructions, descriptions of available tools, and the bug report (e.g., title and description) as input.
It can execute for up to $N$ steps per trajectory.
In each step, the agent takes the current trajectory history, and produces an LLM output and an associated tool execution.
The available tools are: code search, viewing a file, editing a file, running specific tests, and finishing the repair trajectory.
The LLM output consists of (1) a thought in natural language that describes the agent’s reason for its current action and its latest plan for resolving the task, and (2) a function call via Gemini function calling~\cite{comanici2025gemini}.
The LLM output and environment response are appended to the current trajectory history.
For a given bug, we initiate $M$ repair trajectories in parallel to generate at most $M$ patches, and pass them to the patch validation \& selection component. 

The patch validation \& selection component employs a set of patch reviewers to assess patch quality. These reviewers are:
\begin{itemize}
    \item A build \& test reviewer that builds post-patch code, runs tests depending on the patched files, and rejects patches with build errors or test failures.
    \item A smell reviewer that rejects a patch based on simple heuristics, e.g., the last test execution in the trajectory is a failure.
    \item A spec-based reviewer that uses an LLM to generate requirements of the fix (without access to any ground truth) and uses the same LLM to reject patches that do not satisfy the requirements.
\end{itemize}

After filtering out all patches that were rejected by any one of the reviewers, the selector normalizes the remaining patches, mapping each unique likely identifier string to a unique index and normalizing whitespace; groups them by implementation, i.e., patches with the same normalized contents are grouped together; and selects a patch from the largest group (i.e., majority voting to select the most frequently occurring implementation).
For tie-breaking, the selector selects a smaller patch by line count---a simple heuristic to select patches with fewer spurious changes. 

\subsection{Cogeneration of BRT in Agentic APR}
\label{sec:background:motive}

Our motivation for cogenerating BRT in agentic APR is mainly driven by feedback and observations from production.
We now describe our main factors identified.

\subsubsection{Feedback from Code Owner}
\label{sec:background:motive:feedback}
Based on discussions and code comments from code owners that review our AI-generated bug fixes, they strongly prefer fix patches to also have an accompanying fail-to-pass BRT. 
Such tests increase their confidence in the quality of the candidate fix, and support them in accepting a patch that adheres to internal coding standards.
This motivation also largely aligns with the importance of BRT discussed in recent work~\cite{cheng2025agentic,mundler2024swt,ahmed2025heterogeneous,kang2025autocodesherpa}. 

\subsubsection{Test Prevalence in Human-Written Fixes}
\label{sec:background:agent:motive:testprevalence}

We observe a prevalence of test changes (in addition to the core fix logic) in human-written patches for bugs in our deployment of Passerine, such as for null pointer exception (NPE) bugs.
With the goal to improve AI-generated NPE patches, we analyze the attributes of human-written NPE patches during 2024, and find that 39\% of the human-written NPE patches include a test change.  

To obtain this metric, we first derive 7 attribute categories by manually inspecting 20 random, human-written NPE patches. These 7 categories are improvements to: null check logic, test code, base case handling, error message, root cause logic, type annotation, and others. A patch can have multiple attributes. We then run an LLM annotator on the same samples, and verify the LLM annotations reach substantial agreement with the manual annotations, i.e., the Fleiss’s Kappa ranges from 0.61 to 0.81 across categories ~\cite{fleiss1971measuring}. 

We apply the LLM annotator to all human-written NPE patches. From these annotations, we observe that the most common attribute is adding null checks (69\%), and the second most common attribute is test changes (39\%), i.e., adding a new test or modifying an existing test to reproduce the NPE.
This observation motivates us to cogenerate a BRT for the given fix in, ideally, every patch.

\section{Dynamic Cogeneration}
\label{sec:impl}

\begin{figure*}[t]
    \centering
    \includegraphics[width=\textwidth]{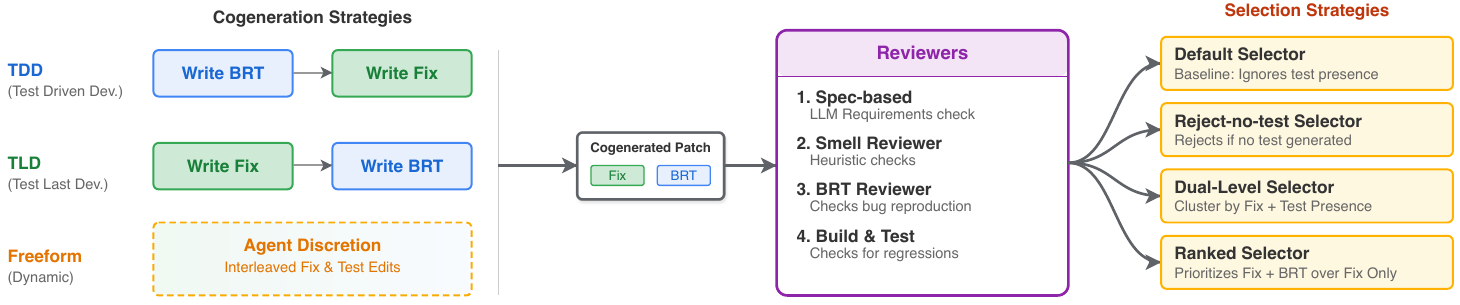}
    \caption{Overview of our agentic APR system under cogeneration. The APR agent utilizes one of three cogeneration strategies: (1) Test-Driven Development (TDD), (2) Test-Last Development (TLD), or (3) Freeform. This process cogenerates patches containing both a code fix and a Bug Reproduction Test (BRT). These patches undergo a multi-stage validation process by a suite of reviewers (spec-based, smell, BRT, and build \& test) to assess validity and reproducibility. Finally, a patch selection strategy is used to select the most promising patch among the candidates.}
    \label{fig:overview}
\end{figure*}

Figure~\ref{fig:overview} provides an overview of our agentic APR system under cogeneration.
The process begins with the APR agent employing one of three \textit{Cogeneration Strategies}: TDD, TLD, or Freeform (\S\ref{sec:impl:cogen}).
Regardless of the strategy, the agent produces \textit{Cogenerated Patches} containing both a fix and a BRT.
These patches are then assessed by a suite of \textit{Reviewers}, which includes checks for bug reproduction (\S\ref{sec:impl:brtreviewer}), regression testing, and specification compliance.
Finally, valid candidates proceed to the \textit{Selection Strategies} phase, where an optimal patch is chosen based on criteria such as the presence of a candidate BRT (\S\ref{sec:impl:selection}).

\subsection{Implementation of Cogeneration}
\label{sec:impl:cogen}
To enable cogeneration, we add the text in Listing~\ref{cogen_prompt} to the agent’s system instructions, which instructs the agent to implement and return the fix and the BRT in the same trajectory.

\begin{listing}[t!]
\centering
\caption{Additional system instruction for cogeneration.}
\label{cogen_prompt}
\begin{lstlisting}
If the bug report mentioned a reproduction test, you will use it to reproduce the bug and verify your fix. If the bug report did not mention a reproduction test, that means no existing test reproduces the bug. Then your plan should include writing such a test that reliably reproduces the bug. You need to find the most relevant test file to write a new test.
\end{lstlisting}
\vspace{-10pt}
\end{listing}

\begin{listing}[t!]
\centering
\caption{Additional TDD instruction.}
\label{tdd_cogen_prompt}
\begin{lstlisting}
You must first write a reproduction test before you can write the fix for the bug. Run the reproduction test to confirm it reliably reproduces the bug before you write the fix. After you have written the fix, run the reproduction test again to verify your fix resolves the bug. Refine your fix until the reproduction test passes.
\end{lstlisting}
\vspace{-10pt}
\end{listing}

\begin{listing}[t!]
\centering
\caption{Additional TLD instruction.}
\label{tld_cogen_prompt}
\begin{lstlisting}
You must first write the fix for the bug before you can write a new reproduction test. Run the reproduction test to verify your fix resolves the bug. Refine your fix until the reproduction test passes.
\end{lstlisting}
\vspace{-10pt}
\end{listing}

In addition, we implement and evaluate a cogeneration variant that practices Test-Driven Development (TDD)~\cite{beck2003test,munir2014experimental,janzen2005test,fucci2016dissection}, by further appending the text in Listing~\ref{tdd_cogen_prompt}.
TDD cogeneration instructs the agent to first implement the BRT and verify its bug reproducibility before implementing the fix.
The assumed benefit of TDD is that the BRT could provide a better understanding of the root cause and lead to a higher-quality fix~\cite{gao2025trae,yang2024swe,ahmed2024tdd,beck2003test}.

We also implement and evaluate a variant that practices Test-Last Development (TLD)~\cite{munir2014experimental,beller2015and,fucci2016dissection,zhou2026change}, by appending the text in Listing~\ref{tld_cogen_prompt}.
TLD cogeneration instructs the agent to first implement the fix before it can add a BRT to verify the fix to strengthen the test suite.
The expected benefit of TLD is that the comparison between the fix and the bug could facilitate a smoother BRT implementation~\cite{kitsios2025automated}.

\subsection{Validation of BRT in Cogenerated Patch}
\label{sec:impl:brtreviewer}

We implement a reviewer to assess BRT quality, which will be used in patch selection when selecting cogenerated patches.

The BRT reviewer extracts the generated BRT from the patch, and runs it on the buggy code to evaluate whether it can successfully reproduce the bug.
It rejects the patch if the BRT does not fail (i.e., a BRT is expected to fail on the buggy code); it performs no operation (no-op) if the patch is empty. It can also incorporate an LLM-as-a-Judge reviewer~\cite{shi2025towards} to determine whether the BRT failure reproduces the bug, given the bug report, test log, and the patch as the LLM input. The BRT is then run on the cogenerated patch by the build \& test reviewer, which already exists in our APR system, to evaluate whether it passes on the cogenerated fix.


\subsection{Test-Aware Patch Selection}
\label{sec:impl:selection}

The need for developing a test-aware patch selector is driven by the limitation of our original fix-only patch selector in selecting patches with tests, and the simultaneous presence of patches with and without tests under cogeneration.

The patch selector described in \S~\ref{sec:background:agent} was designed for fix-only patches, and frequently discriminates against patches with tests. Because the test code varies more than the fix code, patches containing tests are split into smaller clusters during majority voting, losing out to the larger groups of test-free patches. Furthermore, our current tie-breaking criterion—which favors shorter code—consistently penalizes patches that include necessary BRTs.

Additionally, due to the stochastic nature of LLMs and the agentic workflow, not every patch the APR agent cogenerated contains both non-test and test changes (e.g., which can be related to LLM configuration, bug difficulty, agent decision, etc).
A patch selector should assume and account for that each patch may have only a fix, a BRT, or both.
In production, we prioritize selecting a final patch with both a valid fix and a valid BRT; if no such patch is available, we still want to select a final patch that has a valid fix without a test; otherwise, we do not select any patches.

Accordingly, we implement two test-aware patch selectors:

\textbf{Dual-level test-aware patch selector.}~~
Compared to the default patch selector, this selector groups patches by the tuple $\langle$hash of the fix implementation, whether the patch has test changes$\rangle$, where fix implementation is the normalized content of the non-test files.

In this selector, patches with the same fix implementation may be put into two different groups if some have a test and some do not.
A patch with a test is selected only if the majority has the same fix and a test.
If patches with tests are not the majority, this strategy will select a patch that has no test.

\textbf{Ranked test-aware patch selector.}~~
Compared to the default, this patch selector first groups patches by the hash of the fix implementation. It then ranks patches in each group lexicographically by: whether the patch has test changes (patches with test changes are prioritized), then by the line count after applying the entire patch (patches with fewer lines are prioritized).

In this selector, a patch that shares the same fix implementation as the majority and has a test is selected.

\section{Evaluation Setup}
\label{sec:setup}

We describe the evaluated dataset, research questions, and evaluation setup in each research question.

\subsection{Dataset}
\label{sec:setup:dataset}

We conduct our evaluation on a set of 120 human-reported bugs collected from GITS between May and September 2025.
The dataset curation follows recent studies~\cite{rondon2025evaluating,cambronero2025abstain}, which closely mirrors prior practice~\cite{jimenez2023swe,maddila2025agentic}.
Each bug has a bug report and an associated ground truth patch that contains a fix and a fail-to-pass BRT.
The patch is written by the developer during bug fixing and can be used as a held-out oracle for offline evaluation.
These bugs do not have any BRT prior to their ground truth patches.
When curating bugs for offline evaluation, we do not omit bugs deemed unfit by our online abstention policy (\S\ref{sec:background:agent}), or filter by bug category or difficulty.

Table~\ref{tab:dataset} lists the number of bugs by programming languages.
Additionally, each bug must satisfy the following requirements:

\setlength{\intextsep}{5pt}%
\setlength{\columnsep}{5pt}%

\begin{wraptable}{r}{25mm}
\centering
\caption{Dataset.}
\label{tab:dataset}
\vspace{-10pt}
  \begin{adjustbox}{width=0.25\columnwidth}
\begin{tabular}{|l|r|}
\hline
\textbf{PL} & \textbf{\# Bugs} \\ \hline
C++ & 36 \\ \hline
Dart & 2 \\ \hline
Go & 17 \\ \hline
Java & 34 \\ \hline
Kotlin & 14 \\ \hline
Python & 17 \\ \hline
\end{tabular}
\end{adjustbox}
\end{wraptable}

\textbf{Bug report.}~~The bug report is text-only and has no attached multimedia, e.g., screenshots.
We exclude changes that affect CSS, HTML, SQL, binary data files, configuration languages, or multimedia data files.
Each bug is manually verified to contain no sensitive personal information.

\textbf{Ground truth patch.}~~The bug and the ground truth patch have a 1:1 correspondence, meaning that the resolution was delivered as a single code change, for ease of comparison between the generated and ground truth patches.
The patch is less than 150 unidiff lines, and had at least 31 test targets that depended on the changed files (empirically determined to prevent expensive test reviews).
The BRT must execute without any build errors; it must fail on pre-patch code and pass on the post-patch code. 

\subsection{Research Questions}
\label{sec:setup:rq}

We aim to study the following research questions:
\begin{itemize}
    \item \textbf{RQ1:} How effective is cogeneration compared to generating a fix alone or BRT alone?
    \item \textbf{RQ2:} How do different cogeneration strategies impact the APR agent's behaviors and expected efficacy?
    \item \textbf{RQ3:} How well do different patch selection strategies identify patches with both fix and BRT under cogeneration?
\end{itemize}

\subsubsection{RQ1: Effectiveness of Cogeneration}
\label{sec:setup:rq:rq1}

Cogeneration instructs the APR agent to simultaneously produce two highly correlated artifacts in bug fixing: a BRT and a fix. However, cogeneration is beneficial only if it allows the agent to produce as many fixes as Fix-only generation, and as many BRTs as BRT-only generation; otherwise, using dedicated agent pipelines for each of these would be more effective. Thus, in this RQ, we compare the fix and BRT generation success rates of our APR agent under cogeneration versus Fix-only and BRT-only.

\paragraph{Evaluated configurations}
We evaluate 5 prompt configurations below.
In each configuration, the agent is also instructed to first analyze the root cause and write a plan before editing any file, and that the plan should be updated as it makes new observations.

Baseline configurations:
\begin{itemize}
    \item \textbf{BRT-only.} The agent is instructed to implement a BRT.
    \item \textbf{Fix-only.} The agent is instructed to implement a fix. 
\end{itemize}

Cogeneration configurations:
\begin{itemize}
    \item \textbf{Freeform cogeneration.} The agent is instructed to implement both a fix and BRT, in no specific order. 
    \item \textbf{TDD cogeneration.} The agent is further instructed to follow TDD, i.e., implementing the test before the fix.
    \item \textbf{TLD cogeneration.} The agent is further instructed to follow TLD, i.e., implementing the test after the fix.
\end{itemize}

For each configuration, we run our APR agent to produce 20 trajectories with a 25-step limit for each evaluated bug.
We use Gemini 2.5 Pro with temperature 0.2 and top-p 0.95 as the LLM~\cite{comanici2025gemini}.
We confirm a comparably low numbers of trajectory exceptions across all evaluated configurations (5.9\%--6.8\%) before further analysis.

\paragraph{Effectiveness analysis}

We use the following success@k metrics to measure effectiveness, which is the expectation over the bugs that at least one trajectory in a sample of size $k$ from 20 trajectories satisfies the given success criterion.
success@k metrics such as pass@k are commonly used to evaluate LLM-based code generation~\cite{chen2021evaluating}.

Fix-related metric:
\begin{itemize}
    \item \textbf{pass@k.} Success criterion is having a plausible fix~\cite{xia2022less}, i.e., a fix that passes the oracle BRT.
\end{itemize}

BRT-related metrics:
\begin{itemize}
    \item \textbf{plausibleBRT@k.} Success criterion is having a plausible BRT, i.e., a test that fails on the buggy code and passes on the oracle fix~\cite{kang2023large,mundler2024swt}.
    \item \textbf{candidateBRT@k.} Success criterion is having a candidate BRT, which we define as a test that fails on the buggy code and passes on the generated fix.
\end{itemize}

Metric for both fix and BRT:
\begin{itemize}
    \item \textbf{(pass \& plausibleBRT)@k.} Success criterion is having a plausible BRT and a plausible fix in the same trajectory.  
\end{itemize}

Note that a plausible patch may not be valid, i.e., it may be rejected by a human reviewer.
We assume that generating more plausible patches increases the probability of finding a valid patch~\cite{xia2022less}; thus, we focus on evaluating effectiveness by patch plausibility.

To further understand the performance gain and loss of cogeneration on the evaluated bugs, we inspect the sets of unique bugs addressed by each configuration. A bug is considered ``addressed'' by a configuration if one of its trajectories generated a plausible fix or a plausible BRT for this bug.

\begin{figure*}[t!]
    \centering
    \begin{subfigure}{0.245\textwidth}
        \includegraphics[width=\linewidth]{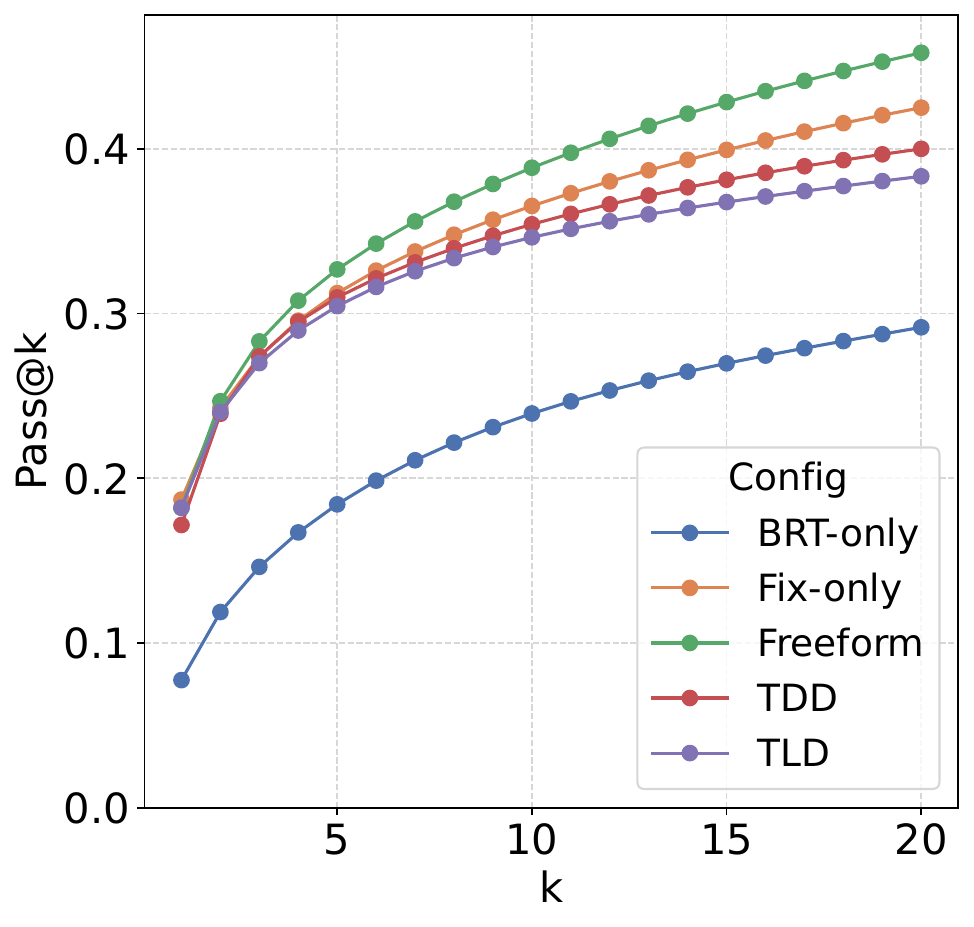}
    \end{subfigure}
    \hfill %
    \begin{subfigure}{0.245\textwidth}
        \includegraphics[width=\linewidth]{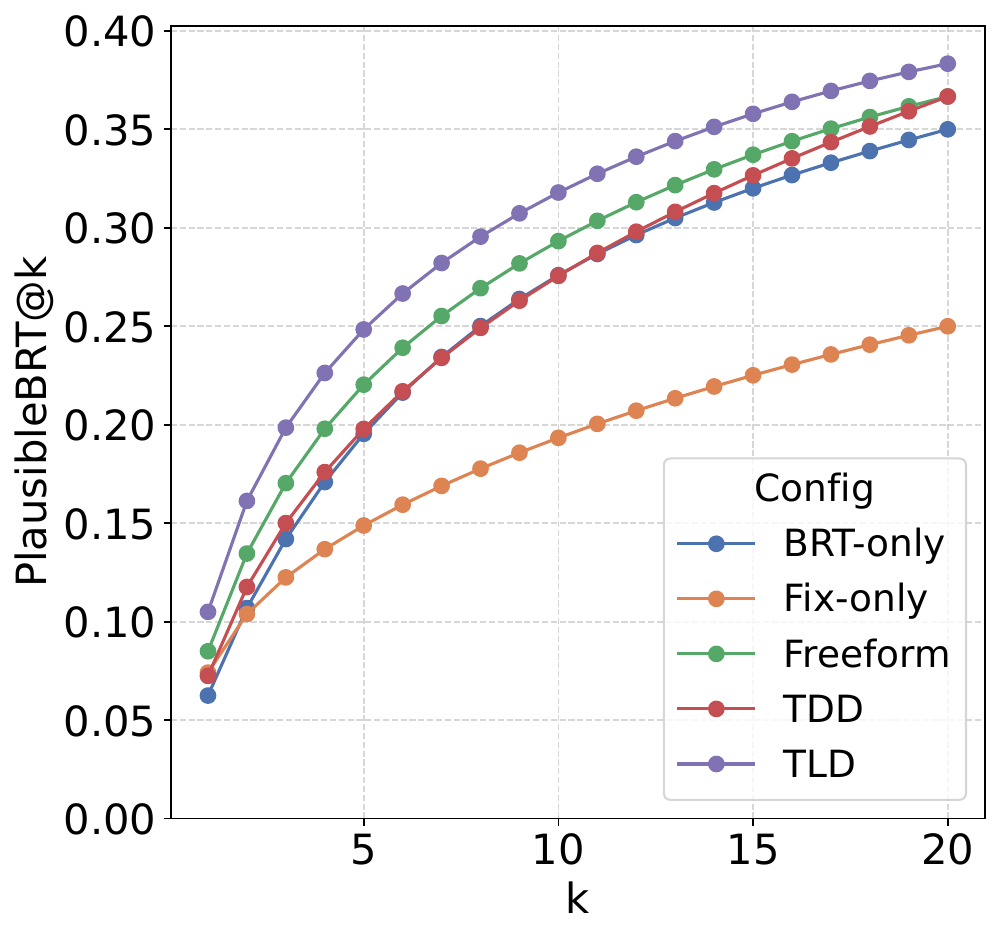}
    \end{subfigure}
    \begin{subfigure}{0.245\textwidth}
        \includegraphics[width=\linewidth]{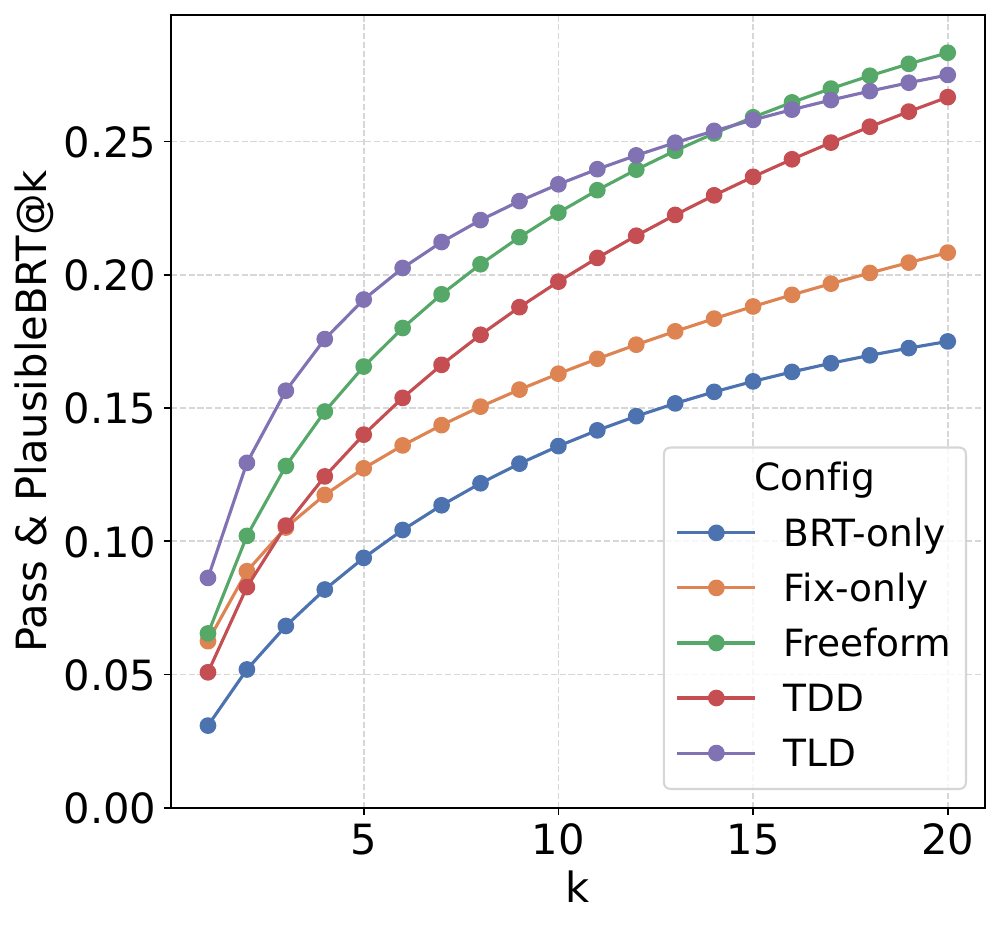}
    \end{subfigure}
    \hfill %
    \begin{subfigure}{0.245\textwidth}
        \includegraphics[width=\linewidth]{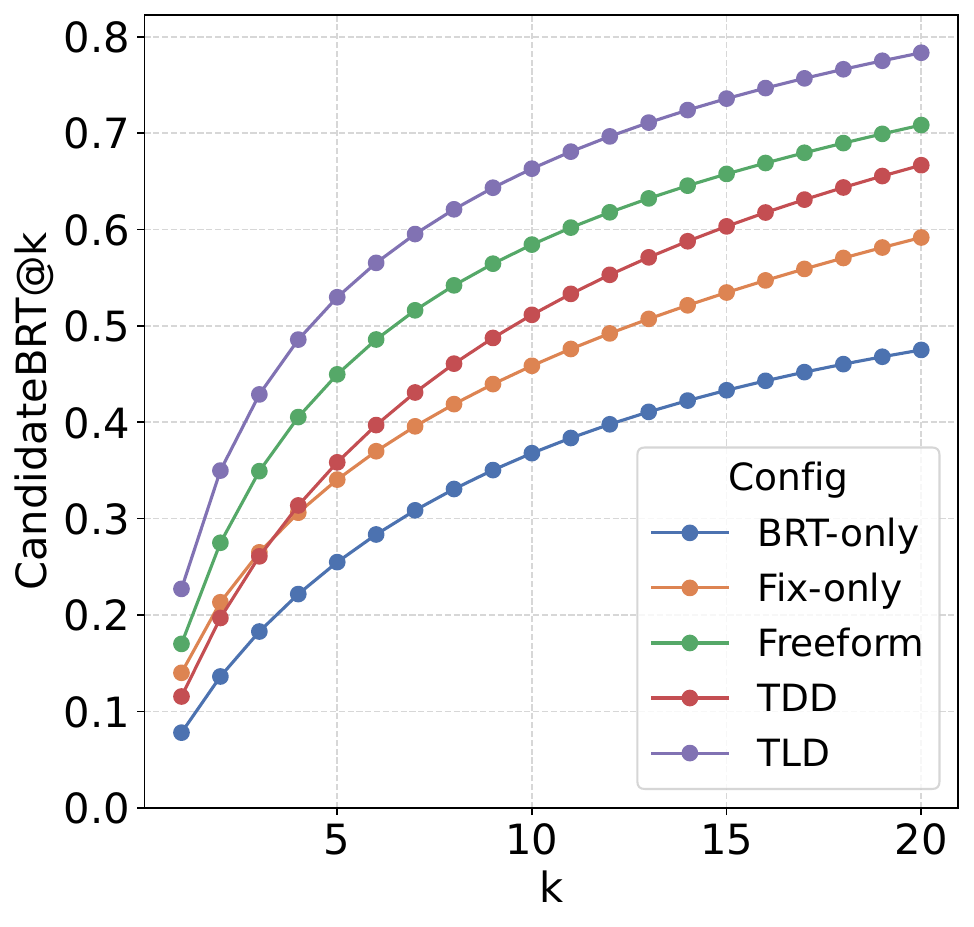}
    \end{subfigure}
    \caption{Effectiveness measured by success@k metrics (\S\ref{sec:setup:rq:rq1}). All cogeneration strategies outperform instructing the agent to generate a fix alone or BRT alone. Freeform cogeneration achieves the highest (pass \& plausible BRT)@20 and pass@20.}
    \label{fig:success_at_k}
\end{figure*}

\paragraph{Efficiency analysis}

To understand overall generation efficiency, we compare the step count distributions of all trajectories.
To understand task resolution efficiency, we compare the step count distributions of the shortest finished trajectory that produces a plausible outcome for each resolved bug.

\subsubsection{RQ2: Characterization of Cogeneration}
\label{sec:setup:rq:rq2}

In this RQ, we investigate how the APR agent’s bug fixing behaviors are influenced by the enforced cogeneration strategies (e.g., TDD versus TLD).
We also define cogeneration success rates to help characterize the expected efficacy of cogeneration.

\paragraph{Agent behavior}

For each configuration, we study the prevalence of the following agent behaviors that are important to bug fixing and cogeneration:

\textbf{Localized edit.}~~Whether the trajectory has a tool call that edits at least one of the files changed by the ground truth patch.
This behavior is important because it signals whether the APR agent effectively attempts to fix the bug.

\textbf{Edited test.}~~Whether the trajectory has a tool call that edits a test file.
We determine test files based on the naming conventions strongly followed at Google.
This behavior is important because it signals whether the agent attempts to reproduce the bug via testing.

\textbf{Edited test first.}~~One major distinction between different cogeneration strategies is when the test is implemented.
To assess whether such distinction exists, we analyze how often a trajectory edits a test file before it edits a non-test implementation file. 

\textbf{Ran test.}~~Whether the trajectory has a tool call that runs tests. 
This behavior is important because it signals whether the agent attempts bug reproduction or fix validation via testing.

\paragraph{State transition modeling}
To systematically characterize the sequential decision-making process of the agent, we model the trajectories as Markov chains.
We abstract the agent's trajectory into a sequence of four primary states: \CodeIn{Start}, \CodeIn{End}, \CodeIn{edit\_test}, and \CodeIn{edit\_source}. We calculate the transition probabilities between these states to quantify the likelihood of the agent moving from one distinct activity to another, e.g., \condprob{edit\_source}{edit\_test}.
To capture the dominant workflow patterns, we visualize these chains by filtering for the highest-probability transitions and removing self-loops (i.e., consecutive actions within the same state).

\paragraph{Cogeneration success}

We define cogeneration success rates as the following conditional probabilities.
With these success rates, we aim to capture how the probability of generating a plausible fix or a BRT can vary given the quality of the previous generation step (e.g., generating fix first) prioritized in each cogeneration strategy: 
\begin{itemize}
    \item \condprob{plausible BRT}{candidate BRT}
    \item \condprob{plausible fix}{candidate BRT} 
    \item \condprob{plausible BRT}{plausible fix}
    \item \condprob{plausible fix}{plausible BRT}
\end{itemize}

\subsubsection{RQ3: Patch Selection for Cogeneration}
\label{sec:setup:rq:rq3}

This RQ focuses on revisiting patch selection for APR, in the context of cogeneration.

\paragraph{BRT signals for fix selection}

BRTs have been used for selecting fixes~\cite{xia2025demystifying,kang2025autocodesherpa,cheng2025agentic,chen2025old}.
We assess whether signals of a BRT can also help indicate the quality of the fix from the same cogenerated patch.
Example BRT signals are whether the patch has a test, a candidate BRT, or a plausible BRT.
We study whether a fix is more likely to be plausible given stronger signals.
We also compute the precision, recall, and F-score of using those BRT signals as binary predictors to whether the same patch has a plausible fix.

\paragraph{Test-aware patch selection}

We evaluate 4 patch selectors (\S\ref{sec:impl:selection}) on selecting a cogenerated patch that contains a plausible fix, a plausible BRT, or both: (1) Default test-unaware, (2) Dual-level test-aware, (3) Ranked test-aware, and (4) Reject-no-test.

All 4 selectors use the same set of reviewers, and additionally the BRT reviewer, which runs BRTs on the bug and rejects patches with non-passing BRT outcomes (\S\ref{sec:impl:brtreviewer}).
The spec-based reviewer uses Gemini 2.5 Pro with temperature 0.0 and top-p 0.95~\cite{comanici2025gemini}.
Compared to the default, the reject-no-test selector rejects all patches that have no test changes.
Similar to our production setting, each patch selector selects at most  1 patch per bug for human review. We evaluate each patch selector on 2,400 patches (120 bugs $\times$ 20 trajectories per bug) from Freeform cogeneration. 

We compute the precision and recall of each patch selector---precision is the percentage of plausible patches over all selected patches; recall is the percentage of bugs with a selected plausible patch over all bugs with at least one plausible patch.
We separately evaluate the three cases where a patch could be plausible: it has a plausible fix, a plausible BRT, or both.

\section{Evaluation Results}
\label{sec:results}

We now present our evaluation results.

\subsection{RQ1: Effectiveness of Cogeneration}
\label{sec:results:rq1}

\subsubsection{Results on success@k} 
Figure~\ref{fig:success_at_k} shows the generation effectiveness results measured by success@k metrics (\S\ref{sec:setup:rq:rq1}). 

Based on pass@k, all cogeneration configurations (i.e., Freeform, TDD, TLD) produce at least one plausible fix for a similar number of bugs compared to Fix-only.
Based on plausibleBRT@k, all cogeneration configurations produce at least one plausible BRT for a similar number of bugs compared to BRT-only.
BRT-only and Fix-only have the worst pass@k and plausibleBRT@k, respectively.

All cogeneration configurations outperform Fix-only and BRT-only on (pass \& plausibleBRT)@k.
Freeform cogeneration, which performs the best, generates at least one patch that has a plausible fix and a plausible BRT for the most number of bugs at $k = 20$. 

Based on candidateBRT@k, TLD cogeneration performs the best.
TLD generates tests that fail on the bug and pass on the generated fix for the most number of bugs, likely because TLD has already written the fix, which can help it write a test that will succeed on that fix and fail on the bug (i.e. the agent has all the states it needs to solve the BRT generation task).

Even when the agent is only instructed to generate a fix or a BRT, the agent may coincidentally also produce a BRT or a fix, respectively, as indicated by the non-zero pass@k achieved by BRT-only and the non-zero plausibleBRT@k achieved by Fix-only.
This observation shows that bug reproduction and bug fixing tasks are highly intertwined and agents exhibit proactive tendencies~\cite{yang2024swe,bouzenia2025understanding}.

Overall, compared to instructing the APR agent to generate only a fix or a BRT, cogeneration allows the agent to produce a fix and BRT in a single patch, without compromising the plausibility rate of either.
Freeform is the most aligned with our goal of prioritizing patches with both fix and BRT over patches with only fix, because it achieves the highest (pass \& plausible BRT)@20 and pass@20.

\subsubsection{Set Analysis on Addressed Bugs}

We now analyze the sets of unique bugs where each configuration generated at least one plausible fix or BRT out of all 20 trajectories per bug on 120 bugs. 

\begin{figure}[t!]
    \centering
    \begin{subfigure}{0.45\columnwidth}
        \includegraphics[width=\columnwidth]{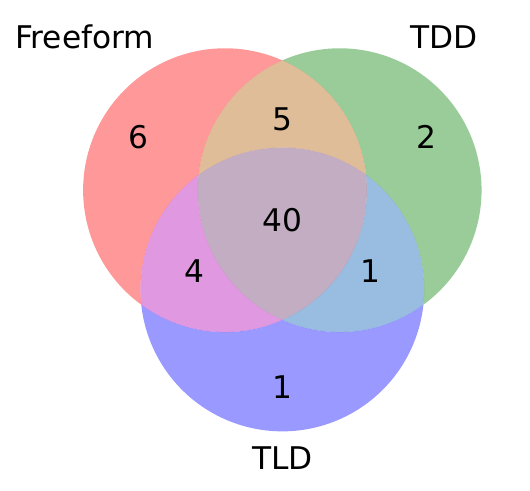}
    \end{subfigure}
    \begin{subfigure}{0.45\columnwidth}
        \includegraphics[width=\columnwidth]{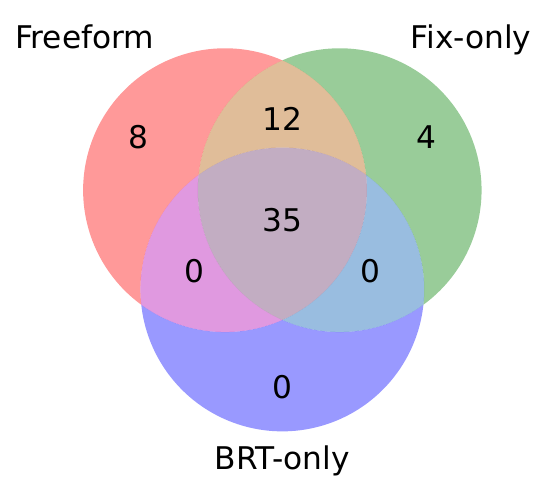}
    \end{subfigure}
    \caption{Venn diagrams of bugs by generated plausible fixes.}
    \label{fig:venn_fix}
\end{figure}

Figure~\ref{fig:venn_fix} compares the set of bugs addressed by each configuration, where ``addressed'' refers to generating a plausible fix.
As shown, Freeform addressed more unique bugs (6 bugs) than TDD (2 bugs) and TLD (1 bug).
Freeform also addressed more unique bugs (8 bugs) than Fix-only (4 bugs) and BRT-only (0 bugs). 

We examine the 4 unique bugs Fix-only addressed but Freeform did not.
Fix-only generated exactly 1 plausible fix for each of the 4 bugs.
We reason that this regression is likely attributed to the randomness in LLM generation.
Meanwhile, on the 8 unique bugs that Freeform addressed but Fix-only did not, Freeform generated exactly 1 plausible fix on 7 bugs and 2 plausible fixes on 1 bug.
This gain could also be attributed to randomness in LLM generation.
Nonetheless, cogeneration still generated plausible fixes for at least as many bugs as Fix-only, while additionally generating BRTs.

\begin{figure}[t!]
    \centering
    \begin{subfigure}{0.45\columnwidth}
        \includegraphics[width=\columnwidth]{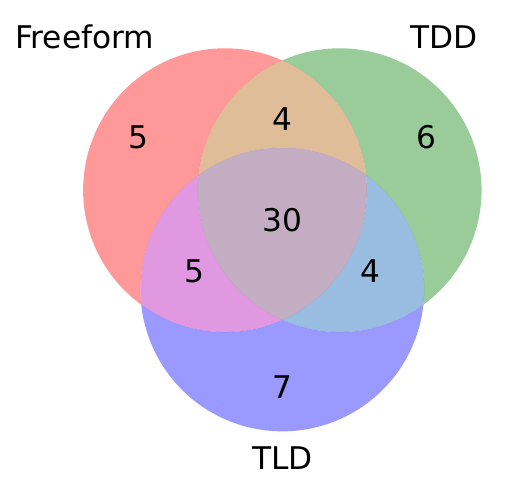}
    \end{subfigure}
    \begin{subfigure}{0.45\columnwidth}
        \includegraphics[width=\columnwidth]{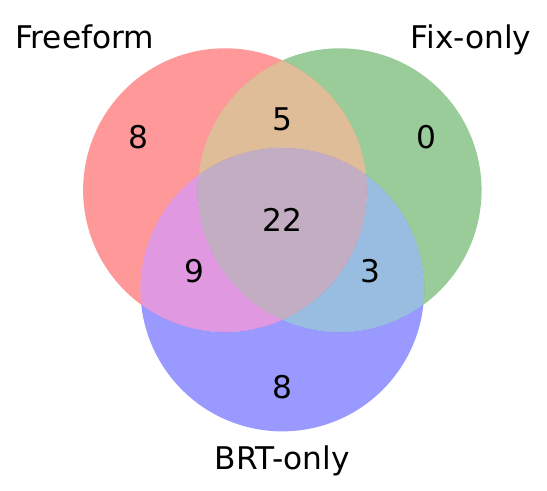}
    \end{subfigure}
    \caption{Venn diagrams of bugs by generated plausible BRTs.}
    \label{fig:venn_brt}
\end{figure}

Figure~\ref{fig:venn_brt} compares the set of bugs addressed by each configuration, where ``addressed'' refers to generating a plausible BRT. 
As shown, all cogeneration strategies addressed a similar number of unique bugs.
Meanwhile, the number of unique bugs addressed by Freeform and BRT-only are the same (8 bugs), both of which are higher than that of Fix-only (0 bugs).

We examine the 8 bugs where BRT-only addressed but Freeform did not, and find that the regression can be attributed to several reasons: randomness in LLM generation given that BRT-only generated exactly one plausible BRT per bug (5 bugs); Freeform generated fewer patches with test, thus reducing the probability of a plausible BRT (2 bugs); tests were generated but not plausible (1 bug).
On the 8 bugs where Freeform addressed but BRT-only did not, the gain can be attributed to: randomness in LLM generation (5 bugs) and many tests were generated but none were plausible (3 bugs).

We also examine the number of addressed bugs by programming languages, but did not notice substantial regression introduced by cogeneration from Fix-only or BRT-only in any specific language.

\subsubsection{Efficiency Results}

Recall that the agent can only execute up to 25 steps per trajectory; we thus use step count to approximate agent efficiency.
Overall, the agent takes more steps towards generating BRTs than towards generating fixes.

\begin{figure}[t!]
    \centering
    \begin{subfigure}{0.48\columnwidth}
        \centering
        \includegraphics[width=\linewidth]{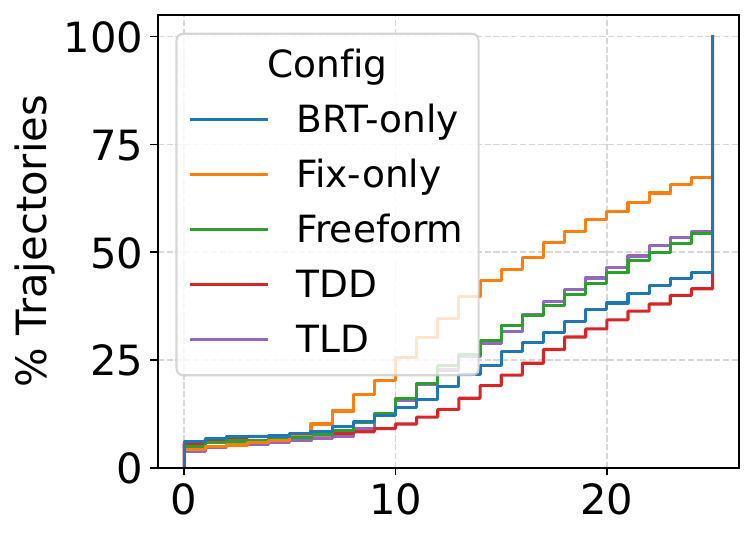}
        \caption{Step count.}
        \label{fig:step_count}
    \end{subfigure}\hfill
    \begin{subfigure}{0.48\columnwidth}
        \centering
        \includegraphics[width=\linewidth]{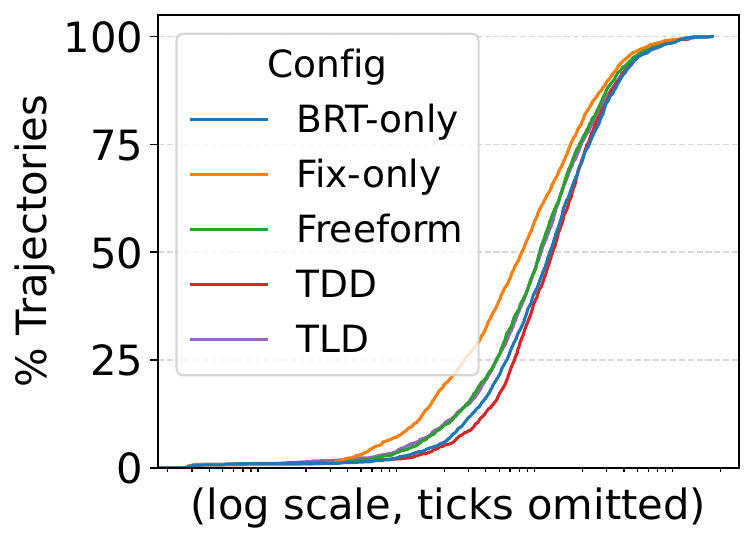} 
        \caption{Token count.}
        \label{fig:token_count}
    \end{subfigure}
    \caption{ECDFs of step count (limit is 25) and token count.}
    \label{fig:step_and_token_count}
\end{figure}

Figure~\ref{fig:step_count} presents the Empirical Cumulative Distribution Function (ECDF) of step counts over all trajectories, where each point $(x, y)$ indicates $y\%$ of the trajectories take at most $x$ steps.
As shown in Figure~\ref{fig:step_count}, we observe that the APR agent takes more steps to finish when it is instructed to prioritize generating a BRT (e.g., BRT-only, TDD) than when it is instructed to prioritize generating a fix (e.g., Fix-only, TLD).
This difference is mostly because BRT generation requires extra steps for running tests, and debugging and repairing unexpected test errors.
In Figure~\ref{fig:token_count}, all cogeneration configurations are as token-efficient as BRT-only, while additionally generating fixes, given the benefit of sharing similar task context.

\begin{figure}[t!]
    \centering
    \includegraphics[width=.7\columnwidth]{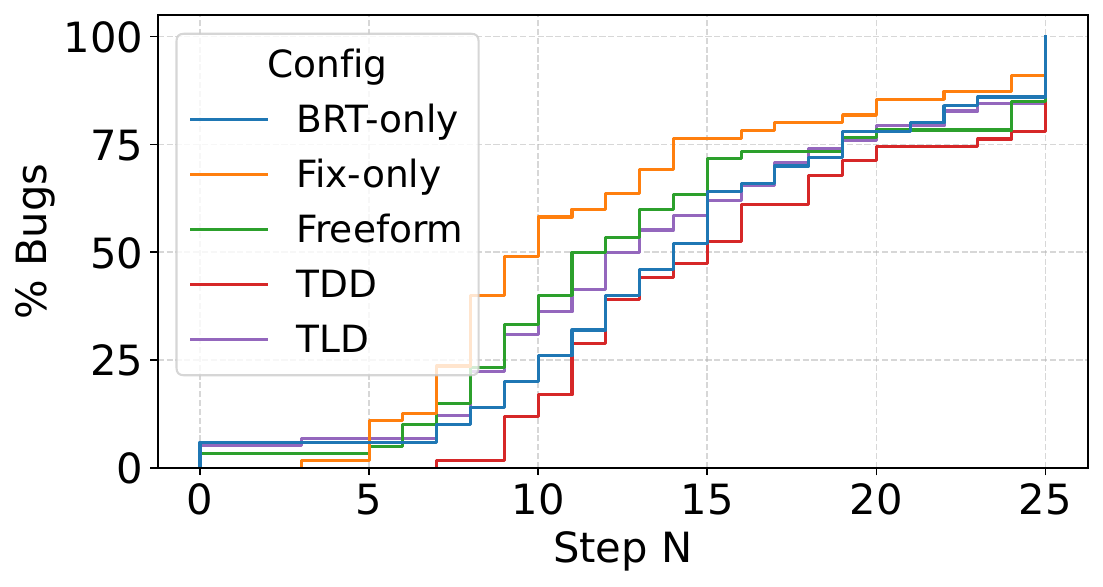}
    \caption{ECDF of step count of the shortest trajectory that produce a plausible outcome in each addressed bug.}
    \label{fig:finish_run_ecdf}
\end{figure}

Figure~\ref{fig:finish_run_ecdf} shows the step count distributions of the shortest trajectory that produces a plausible outcome in each addressed bug, where a plausible outcome refers to the union of all plausible criteria, i.e., Fix-only is successful if the fix is plausible, BRT-only is successful if the BRT is plausible, and cogeneration is successful if at least one of the fix or BRT is plausible.
As shown, there are more bugs with shortest successful trajectories that take fewer steps when the APR agent is not instructed to generate a BRT. 

\subsection{RQ2: Characterization of Cogeneration}
\label{sec:results:rq2}

\subsubsection{Agent Behavior}
\label{sec:results:rq2:behavior}

Table~\ref{tab:behavior_frequency} shows the percentage of trajectories where the agent emitted a specific bug fixing behavior (\S\ref{sec:setup:rq:rq2}).

\paragraph{Successful edit localization}
Across all configurations, the agent successfully edits one or more files changed by the ground truth patch in 39\%--42\% of the trajectories except Fix-only.
The relatively low percentage (31\%) under Fix-only is likely because it often only edits source files (i.e., non-test implementation files).
The agent successfully edits one or more source files changed by the ground truth patch in 22\%--25\% of the trajectories under configurations except BRT-only (14\% under BRT-only). 
These results imply that cogeneration does not substantially affect the localization success rate of the APR agent compared to Fix-only and BRT-only.

\paragraph{Test editing and invocation}
The agent edits test files in 26\%--68\% of the trajectories across all configurations: it edits test files the least frequently under Fix-only, and the most frequently under BRT-only.
Meanwhile, the agent runs tests in 71\%--82\% of the trajectories across all configurations.

When we inspect whether the first file the agent edits is a test file, we find that the agent can largely follow the instructed practices: the APR agent edits test files first in 79\% of the trajectories under TDD, compared to only 4\% under TLD. 

Importantly, under Freeform, the APR agent edits test files first in only 25\% of the trajectories, which implies that it could be biased towards TLD when resolving this set of bugs.
The agent appears to be confident in deriving an initial explanation for a bug, so that it starts with the fix and then tries to confirm the fix with a test.

\begin{table}[t!]
\centering
\caption{\% of trajectories with specific bug fixing behaviors. }
\label{tab:behavior_frequency}
\begin{adjustbox}{width=\columnwidth}
\begin{tabular}{|l|r|r|r|r|}
\hline
\textbf{Configuration} & \multicolumn{1}{l|}{\textbf{\begin{tabular}[c]{@{}l@{}}Localized\\ edit (\%)\end{tabular}}} & \multicolumn{1}{l|}{\textbf{\begin{tabular}[c]{@{}l@{}}Edited\\ test (\%)\end{tabular}}} & \multicolumn{1}{l|}{\textbf{\begin{tabular}[c]{@{}l@{}}Edited\\ test first (\%)\end{tabular}}} & \multicolumn{1}{l|}{\textbf{\begin{tabular}[c]{@{}l@{}}Ran\\ test (\%)\end{tabular}}} \\ \hline
BRT-only & 39.96 & 68.42 & 81.84 & 81.79 \\ \hline
Fix-only & 30.54 & 25.58 & 2.27 & 75.17 \\ \hline
Freeform & 39.29 & 43.79 & 24.54 & 74.04 \\ \hline
TDD & 42.17 & 55.67 & 78.66 & 77.54 \\ \hline
TLD & 40.67 & 54.29 & 4.16 & 70.54 \\ \hline
\end{tabular}
\end{adjustbox}
\end{table}

\subsubsection{State Transitions}
\begin{figure}[t!]
    \centering
    \includegraphics[width=\columnwidth]{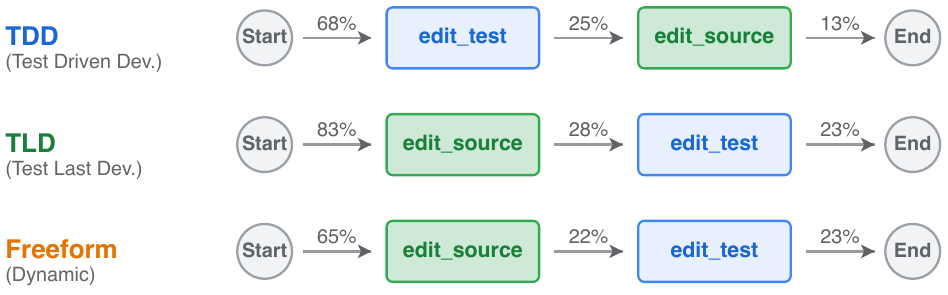}
    \caption{Dominant state transitions. TDD and TLD strictly follow instructed workflows, while Freeform naturally converges to TLD but with lower transition confidence.}
    \label{fig:state_transitions}
\end{figure}

Figure \ref{fig:state_transitions} visualizes the dominant Markov chain transitions.
We observe strict adherence to instructions: the TDD configuration prioritizes \CodeIn{edit\_test} (68\% from \CodeIn{Start}), TLD prioritizes \CodeIn{edit\_source} (83\%).
Note that the calculation of these probabilities from \CodeIn{Start} considers trajectories that have not edited any file, thus these probabilities are different from the probabilities of ``Edited test first'' in Table~\ref{tab:behavior_frequency}, whose calculation only considers trajectories that have edited at least one file.

As shown in Figure \ref{fig:state_transitions}, notably, the Freeform configuration reveals a natural ``Fix First'' bias, mimicking the TLD workflow with a 65\% transition to source editing.
However, this probability is lower than in the enforced TLD configuration, indicating higher behavioral entropy when the workflow is not explicitly constrained.

\subsubsection{Cogeneration Success}

We compute the conditional probabilities in Table~\ref{tab:cogen_success} as cogeneration success rates (\S\ref{sec:setup:rq:rq2}) on trajectories where the agent willingly finishes (i.e., invokes the finish command) before reaching the step limit.

TDD achieves a higher \condprob{plausible\ BRT}{candidate\ BRT} and \condprob{plausible fix}{candidate BRT}, likely due to its advantage on refining the BRT to be more robust and capture the bug more comprehensively.
TLD achieves a higher \condprob{plausible BRT}{plausible fix}, likely due to TLD writing a test after the fix: if the fix is plausible, the test that matches the fix is more likely to be plausible.
All cogeneration strategies achieve similarly high values in \condprob{plausible fix}{plausible BRT}, implying that a cogeneration trajectory is highly likely to have a plausible fix given the presence of a plausible BRT.
Note that plausibility is tested with an oracle, so we cannot take advantage of \condprob{plausible *}{plausible *} but only the \condprob{plausible *}{candidate *} at deployment time (\S\ref{sec:results:rq3:brtsignal}).

\begin{table}[t!]
\centering
\caption{Conditional probabilities on cogeneration success. }
\label{tab:cogen_success}
\begin{adjustbox}{width=\columnwidth}
\begin{tabular}{|l|r|r|r|}
\hline
\textbf{Conditional probability} & \textbf{Freeform} & \textbf{TDD} & \textbf{TLD} \\ \hline
\condprob{plausible BRT}{candidate BRT} & 0.45 & 0.53 & 0.44 \\ \hline
\condprob{plausible fix}{candidate BRT} & 0.45 & 0.53 & 0.45 \\ \hline
\condprob{plausible fix}{plausible BRT} & 0.86 & 0.89 & 0.87 \\ \hline
\condprob{plausible BRT}{plausible fix} & 0.39 & 0.33 & 0.54 \\ \hline
\end{tabular}
\end{adjustbox}
\end{table}

\subsection{RQ3: Patch Selection for Cogeneration}
\label{sec:results:rq3}

\subsubsection{BRT Signals for Fix Selection}
\label{sec:results:rq3:brtsignal}

Table~\ref{tab:brt_signal_prob} presents the conditional probability that a cogenerated patch has a plausible fix given that it emits a certain BRT signal, calculated on all cogeneration trajectories.
Compared to selecting from all cogenerated patches (i.e., ``No filter''), it is 2$\times$ more likely to select a patch that has also a plausible fix if we select from those with a candidate BRT; if we consider selecting from cogenerated patches that have a plausible BRT, it is 4$\times$ more likely to select a patch with a plausible fix. 

At deployment time, the plausibility of a BRT is unknown because its assertion requires an oracle fix, thus this value is a potential upper bound of the BRT's utility for fix validation.
Nonetheless, we can still obtain useful signals from the BRT with more comprehensive assessment at deployment time.
For example, we can use an LLM reviewer (which takes the bug report, patch, and test log as input) and determine whether the candidate BRT’s failure is related to the bug.
On the other hand, we can also use an LLM judge (which additionally takes ground truth patch as input compared to LLM reviewer)~\cite{shi2025towards} to estimate a more accurate upper bound of BRT utility for offline APR evaluation.

\begin{table}[t!]
\centering
\caption{Conditional probabilities of a plausible fix in the patch given certain BRT signals. ``*'' signals require oracle.}
\label{tab:brt_signal_prob}
\begin{adjustbox}{width=0.9\columnwidth}
\begin{tabular}{|l|r|r|r|}
\hline
\textbf{BRT signal as patch filter} & \multicolumn{1}{l|}{\textbf{Freeform}} & \multicolumn{1}{l|}{\textbf{TDD}} & \multicolumn{1}{l|}{\textbf{TLD}} \\ \hline
No filter (baseline) & 0.18 & 0.17 & 0.18 \\ \hline
No test (baseline) & 0.14 & 0.19 & 0.11 \\ \hline
Test & 0.24 & 0.16 & 0.24 \\ \hline
CandidateBRT & 0.44 & 0.48 & 0.43 \\ \hline
LLM-reviewed CandidateBRT & 0.45 & 0.50 & 0.45 \\ \hline
PlausibleBRT* & 0.77 & 0.70 & 0.82 \\ \hline
LLM-judged PlausibleBRT* & 0.78 & 0.71 & 0.85 \\ \hline
\end{tabular}
\end{adjustbox}
\end{table}

\begin{figure}[t!]
    \centering
    \includegraphics[width=\columnwidth]{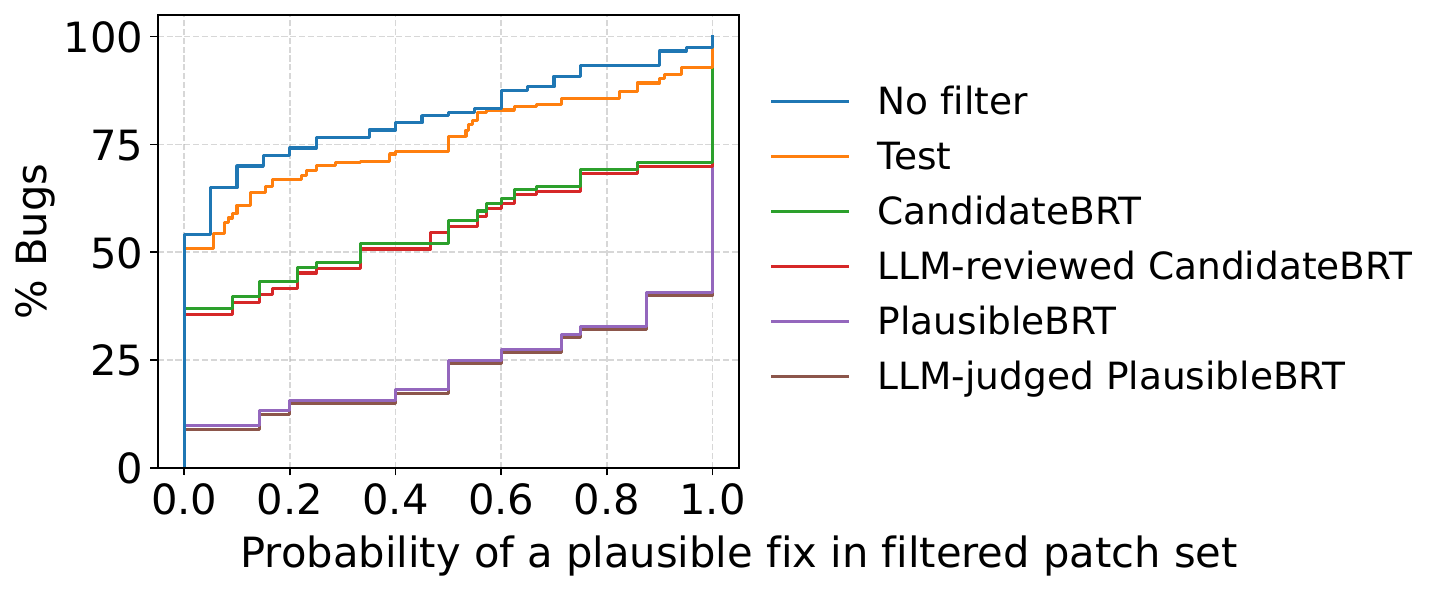}
    \caption{ECDF of conditional probability of a patch having a plausible fix given it has certain BRT signals among all patches of each bug, on Freeform cogeneration trajectories.}
    \label{fig:brt_signal_bug_ecdf}
\end{figure}

Figure~\ref{fig:brt_signal_bug_ecdf} presents the distribution of bugs by these conditional probabilities on Freeform cogeneration trajectories. It further confirms that, within all the cogenerated patches of each bug, selecting a patch with certain BRT signals can improve the likelihood of selecting a patch that has a plausible fix.
This observation also holds for other cogeneration trajectories.

Table~\ref{tab:brt_signal_predictor} shows the performance of using these signals to predict whether the corresponding patch has a plausible fix on Freeform cogeneration.
We can observe that the precision increases and recall decreases as we impose more stringent quality signals on the BRT.
This observation also holds on other cogeneration trajectories.

\begin{table}[t!]
\centering
\caption{Performance of using BRT signals as fix plausibility predictors on Freeform cogeneration trajectories.}
\label{tab:brt_signal_predictor}
\begin{adjustbox}{width=\columnwidth}
\begin{tabular}{|l|r|r|r|}
\hline
\textbf{BRT signal} & \textbf{F-score} & \textbf{Precision} & \textbf{Recall} \\ \hline
Test & 0.34 & 0.24 & 0.58 \\ \hline
CandidateBRT & 0.43 & 0.44 & 0.41 \\ \hline
LLM-reviewed CandidateBRT & 0.43 & 0.45 & 0.41 \\ \hline
PlausibleBRT* & 0.49 & 0.77 & 0.36 \\ \hline
LLM-judged PlausibleBRT* & 0.49 & 0.78 & 0.36 \\ \hline
\end{tabular}
\end{adjustbox}
\end{table}

\subsubsection{Test-Aware Patch Selection}

Table~\ref{tab:selector_result} shows the performance of different patch selectors for selecting patches from Freeform cogeneration trajectories (\S\ref{sec:setup:rq:rq3}).
We compute the precisions and recalls on selecting patches that have plausible fixes and/or plausible BRTs. Recall that in production, our priority is selecting a patch that have both; if such a patch does not exist, we prioritize selecting patches that only have a plausible fix (as opposed to the ones with only plausible BRTs); otherwise, we prefer not to select a patch.

A preferred selector should achieve a recall on patches with plausible fixes as good as the default selector, i.e., not increasing the risk of missing out on patches with high-quality fixes. Given this condition, the selector should also achieve the best precision out of all selectors in selecting patches that also have plausible BRTs.

Based on Table~\ref{tab:selector_result}, all selectors have similar precisions on plausible fix compared to the default baseline (i.e., 0.30--0.32), while the ranked and dual-level selectors have relatively close recall on plausible fix against the baseline.
When we focus on the precision and recall on selecting patches with both plausible fix and BRT, the ranked selector outperforms the default and two-level selector with the higher precision and recall. 

While the reject-no-test selector, which rejects any patches with no test change, achieves high precision and recall on patches with both plausible fix and BRT, it unfortunately has a low recall on plausible fixes, thus missing out on more patches with only plausible fixes. 
The default selector (\S\ref{sec:background:agent}), which ignores test changes, achieves the lowest precision/recall on selecting patches with both plausible fix and BRT. 

Overall, the ranked selector would be preferred in production since it is the most aligned with our objective.

\begin{table}[t!]
\centering
\caption{Precision/recall of selectors on patches with plausible fix and/or BRT on Freeform cogeneration trajectories.}
\label{tab:selector_result}
\begin{adjustbox}{width=\columnwidth}
\begin{tabular}{|l|rr|rr|rr|}
\hline
\multirow{2}{*}{\textbf{Selector}} & \multicolumn{2}{c|}{\textbf{Plausible fix}} & \multicolumn{2}{c|}{\textbf{Plausible BRT}} & \multicolumn{2}{c|}{\textbf{Both}} \\ \cline{2-7} 
 & \multicolumn{1}{c|}{\textbf{Prec.}} & \multicolumn{1}{c|}{\textbf{Rec.}} & \multicolumn{1}{c|}{\textbf{Prec.}} & \multicolumn{1}{c|}{\textbf{Rec.}} & \multicolumn{1}{c|}{\textbf{Prec.}} & \multicolumn{1}{c|}{\textbf{Rec.}} \\ \hline
Default & \multicolumn{1}{r|}{0.30} & 0.81 & \multicolumn{1}{r|}{0.11} & 0.61 & \multicolumn{1}{r|}{0.08} & 0.57 \\ \hline
Dual-level & \multicolumn{1}{r|}{0.32} & 0.79 & \multicolumn{1}{r|}{0.12} & 0.52 & \multicolumn{1}{r|}{0.10} & 0.60 \\ \hline
Ranked & \multicolumn{1}{r|}{0.31} & 0.76 & \multicolumn{1}{r|}{0.19} & 0.69 & \multicolumn{1}{r|}{0.16} & 0.71 \\ \hline
Reject-no-test & \multicolumn{1}{r|}{0.32} & 0.54 & \multicolumn{1}{r|}{0.33} & 0.66 & \multicolumn{1}{r|}{0.24} & 0.71 \\ \hline
\end{tabular}
\end{adjustbox}
\end{table}

\section{Discussion on Cogeneration Failures}
\label{sec:discussion}

We perform manual inspection and LLM-based trajectory analysis to understand common failures that occur in cogeneration, and to discuss mitigation strategies.
We only consider trajectories that have edited the ground truth files but produced undesired outcomes in the 4 scenarios described in Table~\ref{tab:failure}.
We define a failure category by the values of the tuple of these 4 scenarios where at least one outcome is undesired.
For example, $\langle$Pass, No test, No test, No test$\rangle$ would be considered one failure category, and $\langle$Not pass, Fail, Not pass, Not pass$\rangle$ would be another category. 
We then count the number of trajectories and bugs per category.

\begin{table}[t!]
\centering
\caption{Outcomes of a cogeneration trajectory. }
\label{tab:failure}
\begin{adjustbox}{width=\columnwidth}
\begin{tabular}{|l|l|l|}
\hline
\textbf{Test execution scenario} & \textbf{Desired} & \textbf{Undesired} \\ \hline
Oracle BRT on generated fix & Pass & Not pass \\ \hline
Generated BRT on bug & Fail & Not fail, or no test \\ \hline
Generated BRT on generated fix & Pass & Not pass, or no test \\ \hline
Generated BRT on oracle fix & Pass & Not pass, or no test \\ \hline
\end{tabular}
\end{adjustbox}
\end{table}

We find that the top-5 most frequent failure categories are the same across all cogeneration configurations.
In each configuration, they constitute 74\%--79\% of all failed trajectories; the 5th category is at least 3$\times$ larger than the 6th category.
We randomly sample and inspect 10\% of the trajectories per top-5 category from Freeform cogeneration (i.e., 63 inspected trajectories). 

We summarize the patterns of failure and root cause below:

\textbf{No test change.}~~
The primary cogeneration failure is that a trajectory produces a patch with no test.
The main root cause is that the agent had indeed written a test but cleaned up the test changes before finishing, because (1) the BRT was treated as a temporary change to verify the fix (77\%), or (2) the agent could not fix errors in the broken test and performed a clean-up (23\%).
Sometimes the agent also believes adding a BRT is not needed, because it assumes (1) an existing test can reproduce the bug, or (2) only checking regression on existing tests is needed.
An easy way to mitigate these issues is improving the agent’s system instructions. 

\textbf{Debugging challenges.}~~
Another common failure is the agent reaching the step limit before finishing its implementation.
The agent got stuck in a debugging loop, which can be due to different reasons, e.g., being sidetracked by errors of broken tests, finding similar implementations for code understanding, or incorrect tool calls~\cite{bouzenia2025understanding}.
The agent may be tempted to pursue a spurious hypothesis under cogeneration, such as revising tests to make them pass on an actually broken fix, or abandoning a correct fix after observing an unrelated/flaky test error.
Frequent revisions sometimes trap the agent in a polluted workspace, and make it more prone to misdiagnosis.
These issues also occur in general APR agents, which could be mitigated by designing more intelligent agent tools.

\textbf{Fix-BRT interference.}~~
The agent sometimes implements a fix biased towards its own BRT, or vice versa.
The fix fails on the oracle BRT, because it did not address the bug comprehensively (e.g., accounting for all corner cases) due to root cause misdiagnosis, or revisions over-fitted to the generated BRT whose implementation is incomprehensive or different from the oracle BRT.
A test that fails on the oracle fix is sometimes implemented after the generated fix (which can be incorrect or implemented differently than the oracle fix), and the passing outcome stops the agent from scrutinizing its implementations.
Note that overfitting occurs in APR even when the BRT is from external sources~\cite{smith2015cure}.
To mitigate these issues in online APR deployment, we should apply stringent patch validation. 

\paragraph{LLM-based trajectory analysis}

We also develop an LLM-based analysis framework to generate inspection reports for APR agent trajectories.
The framework generates one report per trajectory given the agent prompt, trajectory, and evaluation results.
Each report aims to assess tool usage, instruction following, task completion, patch quality, infrastructure errors, and more.
It then generates an aggregated summary from all the reports. 

We apply this framework to all failed Freeform cogeneration trajectories, after confirming that the LLM-generated observations are largely similar to the manual inspections on the manually inspected ones.
The aggregated LLM observations are similar to our manual inspection results, and additionally identify that tool frictions, such as troubles in finding the right test target when the target names are different from the changed test file name, can contribute to suboptimal cogeneration outcomes. 

\section{Related Work}

Many APR systems generate BRT and mainly use them to validate fixes~\cite{kang2025autocodesherpa,xia2025demystifying,ruan2025specrover,arora2024masai,gao2025trae,yang2024swe,li2025infcode}.
Modular APR systems with deterministic workflows~\cite{xia2025demystifying,ruan2025specrover,arora2024masai} generate BRTs and fixes independently in separate components, run the fixes on the BRTs and return the passing fixes as final output.
Some APR systems~\cite{yang2024swe,gao2025trae} use TDD-style prompts to generate a reproduction script as intermediate code changes for fix validation, and return only the fix as final output. 
Like prior work, we leverage the generated BRT to assess fix quality.
Complementarily, we identify factors in our APR deployment that support cogeneration of fix and BRT, we study the impact of different cogeneration strategies, analyze trajectory dynamics, and present end-to-end results on patch selection under cogeneration.

BRT generation has also gained more interest as BRTs are being increasingly used to assess AI-generated fixes.
Recently, researchers have proposed a number of LLM-based BRT generation techniques, e.g., inverting LLM-generated assertions~\cite{khatib2025assertflip}, combining with search-based software testing~\cite{kitsios2025automated}, improving bug reproducibility with feedback~\cite{nashid2025issue2test}, planning~\cite{ahmed2025otter}, mutating bug reports and using generated fix for BRT selection~\cite{ahmed2025heterogeneous}, or LLM training for BRT generation~\cite{soni2026swe}.
While these approaches follow a deterministic workflow consisting of task-specific stages, there are also agentic BRT generators~\cite{mundler2024swt,cheng2025agentic}. 
We focus on instructing an APR agent to cogenerate a BRT along with the fix in the same trajectory, rather than building a dedicated BRT or fix generator.

\section{Threats to Validity}

\textbf{Internal.}~~
Our dataset curation closely follows prior work~\cite{jimenez2023swe,chen2021evaluating,maddila2025agentic}, which could still introduce biases towards resolvable issues, potentially inflating the agent's success rate.
Our manual assessments (\S\ref{sec:background:motive:feedback}, \S\ref{sec:discussion}) may diverge from those of the actual code owners, and contain subjective interpretations on complex failures.
We aimed to mitigate these threats via LLM-as-a-Judge~\cite{zheng2023judging}.

We use a temperature 0.2 to encourage stability while balancing generation diversity, however, LLM sampling noise can still affect agent performance (\S\ref{sec:results:rq1}), which we mitigate by running 20 trajectories per bug per configuration (\ref{sec:setup:rq:rq1}).

\noindent\textbf{External.}~~
We conduct our study in Google’s proprietary engineering environment, which reflects specific internal practices (e.g., monorepo dependencies).
Impact from cogeneration may differ on environments that adhere to substantially different practices.

Since agent performance often depends on the underlying LLM, our findings regarding regarding different cogeneration strategies could shift as new LLMs become available.

\noindent\textbf{Construct.}~~
Following the de facto standards in APR~\cite{yang2024swe,chen2021evaluating}, we measure effectiveness primarily with execution-based metrics (e.g., pass@k) that use held-out tests (\S\ref{sec:setup:dataset}).
However, patches that pass all held-out validation tests may still contain undesired properties leading to rejection during code owner review~\cite{yu2025utboost,wang2025solved,zhu2025establishing}.
Our reported pass rates thus may overestimate the true validity of the patches compared to a strict human-in-the-loop process.

Our evaluation and patch selection rely partly on LLM-based validators (e.g., the spec-based reviewer) and heuristic checks.
These automated judges may lack the domain context to distinguish between a genuinely correct patch and one that satisfies superficially generated requirements.
If the validator's generated requirements miss case-dependent nuances, it may accept invalid patches or reject valid ones, affecting the measured precision of our approach.

\section{Conclusion}

We present a study of agentic APR to dynamically generate and return to the user both fix and BRT in the same trajectory.
We study the effectiveness of such a cogeneration setup and find that cogeneration allows the APR agent to generate plausible BRTs for as many bugs as a dedicated BRT-only agent that only generates BRTs, and plausible fixes for as many bugs as a Fix-only agent that only generates fixes, thereby reducing engineering effort in maintaining and coordinating separate fix/BRT generation pipelines at scale. 

We also present extensive analyses to understand agentic APR under cogeneration.
These analyses include characterizing the impact of our agent’s actions from different cogeneration strategies, assessing the performance of test-unaware and test-aware patch selection strategies for selecting patches containing both a plausible fix and plausible BRT, and inspecting common reasons why cogeneration trajectories can produce undesired outcomes.
We hope this study complements the existing work on agentic APR, patch validation, patch selection, and BRT generation.

\balance
\bibliographystyle{ACM-Reference-Format}
\bibliography{references}

\end{document}